\documentclass[11pt]{article}
\pdfoutput=1 %
\usepackage[table]{xcolor}
\usepackage{jheppub} 
                     
\usepackage{braket,slashed,bm}
\usepackage{array,multirow}
\usepackage{simplewick}
\usepackage[T1]{fontenc} % if needed
\usepackage[force]{feynmp-auto}

\usepackage[utf8]{inputenc}

\usepackage{tikz}
\usetikzlibrary{arrows,decorations.pathmorphing,backgrounds,positioning,fit,petri,automata,shadows,calendar,mindmap,
decorations.markings,calc}

\usepackage{amsmath}
\usepackage{amssymb}
\usepackage{graphicx}
\usepackage{esint}
\usepackage{wasysym}
\usepackage{array}
\usepackage{comment}
\usepackage{tabu}
\usepackage{relsize}
\usepackage{pifont}

 \newcommand{\nda}{\scriptstyle{\times}}

\newcommand{\cw}{\cellcolor{white}}

\newcommand{\czb}{\cellcolor{zero2}}

\definecolor{zero1}{rgb}{0.35,0.4,0.85}
\definecolor{zero2}{rgb}{0.88,0.88,.88}
\definecolor{zero3}{rgb}{0.85,0.85,0.95}
\definecolor{zero4}{rgb}{1,0.3,0.3}

\newcommand{\Dfbd}{\mathord{\buildrel{\lower3pt\hbox{$\scriptscriptstyle\leftrightarrow$}}\over {D}_{\mu}}}
\def\CO{\mathcal{O}}

\allowdisplaybreaks[0]

\title{Complete One-Loop Matching for a Singlet Scalar in the Standard Model EFT}

\author[a,b]{Minyuan Jiang,}
\author[b,c]{Nathaniel Craig,}
\author[d,c]{Ying-Ying Li,}
\author[b]{Dave Sutherland}

\affiliation[a]{Department of Physics, Nanjing University, Nanjing 210098, P.R.C.}
\affiliation[b]{Department of Physics, University of California, Santa Barbara, CA 93106, USA}
\affiliation[c]{Kavli Institute for Theoretical Physics, Santa Barbara, CA 93106, USA}
\affiliation[d]{Department of Physics, The Hong Kong University of Science and Technology, Clear Water Bay, Kowloon, Hong Kong S.A.R., P.R.C.}

\emailAdd{dg1522023@smail.nju.edu.cn}
\emailAdd{ncraig@physics.ucsb.edu}
\emailAdd{ylict@connect.ust.hk}
\emailAdd{dwsuth@ucsb.edu}

\abstract{We present the results of the first complete one-loop matching calculation between the real singlet scalar extension of the Standard Model and the Standard Model effective field theory (SMEFT) at dimension six. Beyond their immediate relevance to the precision calculation of observables in singlet extensions of the Standard Model, our results illustrate a variety of general features of one-loop matching. We explore the interplay between non-supersymmetric non-renormalization theorems, the logarithmic dependence of Wilson coefficients, and the relevance of mixed diagrams in theories with large scale separation. In addition, we highlight some of the subtleties involved in computing observables at next-to-leading order in SMEFT by mapping our results to the $T$ parameter at one loop. 
}

\begin{document} 
\maketitle

%%%%%%%%%%%%%%%%%%%%%%%

\section{Introduction} \label{sec:intro}

Abundant expectations for physics beyond the Standard Model are being confronted by the lack of evidence (thus far) for direct production of new states at the LHC. This tension motivates the further development of effective field theory techniques to characterize the imprint of partially decoupled new physics on experimental data, particularly in light of the discovery of the Higgs boson. Significant progress along these lines has been made within the framework of the Standard Model Effective Field Theory (SMEFT), which extends the Standard Model with irrelevant operators for which the electroweak symmetry is linearly realized (see e.g.~\cite{Brivio:2017vri} for a recent review). Within this framework, new physics contributions to experimental observables (or their close relatives) are encoded by Wilson coefficients for irrelevant operators, of which there are finitely many at a given operator dimension. In SMEFT at dimension six there are 59 such (baryon-preserving, single-flavor) non-redundant operators, leading to an effective theory of the form
\begin{equation}
\mathcal{L}_{\rm eff} = \mathcal{L}_{\rm SM} +  \frac{1}{M^2} \sum_{i = 1}^{59} c_i \mathcal{O}_i
\end{equation}
where $M$ is the scale of new physics, $c_i$ are the Wilson coefficients, and $\mathcal{O}_i$ are the irrelevant operators in a given non-redundant operator basis. Here we have taken the Wilson coefficients to be classically dimensionless. In order to minimize potentially large logarithms, Wilson coefficients $c_i$ computed at the matching scale $M$ can be evolved to lower scales $\mu < M$ (perhaps where observables are computed) at one loop using the appropriate matrix of anomalous dimensions $\gamma_{ij}$, 
\begin{eqnarray} \label{eq:cmu1}
c_i(\mu) &=& c_i(M) + \frac{1}{16 \pi^2} \sum_j \gamma_{ij} c_j(M) \log \frac{\mu}{M} \ .
\end{eqnarray}
The full dimension-6 SMEFT matrix of anomalous dimensions in a complete, non-redundant basis (the Warsaw basis \cite{Grzadkowski:2010es}) was computed  in \cite{Grojean:2013kd, Jenkins:2013zja, Jenkins:2013wua, Alonso:2013hga}.

The precision achievable in SMEFT continues to develop at a rapid rate. The effectiveness of SMEFT is being advanced on one hand by improving the precision with which Wilson coefficients in SMEFT are mapped onto observables, and on the other hand by improving the precision with which the Wilson coefficients are themselves computed. The former has been the subject of much recent development; see e.g.~\cite{deFlorian:2016spz, Passarino:2016pzb} for reviews. The latter -- which motivates the current work --  is possible when the specific UV completion in question is weakly coupled, so that matching can be performed in perturbation theory at the scale $M$. Although the calculation of observables in this case can in principle be carried out directly in the UV theory, the EFT framework is valuable insofar as it separates the matching calculation from the mapping calculation and facilitates the resummation of large logarithms.

Recently, considerable progress has been made in improving the precision of matching calculations in perturbative UV completions, including the presentation of a complete tree-level dictionary \cite{deBlas:2017xtg} and the development of master formulae at one loop. In particular, the calculation of Wilson coefficients for perturbative matching has been significantly simplified by the advancement of Covariant Derivative Expansion (CDE) methods \cite{Henning:2014wua} building on earlier covariant functional techniques \cite{Gaillard:1985uh, Chan:1986jq, Cheyette:1987qz}. The CDE master formula for degenerate heavy particles presented in \cite{Henning:2014wua} was later generalized to accommodate non-degenerate heavy particles \cite{Drozd:2015rsp}, dubbed the Universal One-Loop Effective Action (UOLEA). As initially formulated, the CDE methods of \cite{Henning:2014wua, Drozd:2015rsp} did not accommodate mixed diagrams in which both heavy and light particles run in the loop \cite{delAguila:2016zcb} (see also \cite{Boggia:2016asg}), leading to a variety of improvements aimed at capturing these contributions \cite{Henning:2016lyp,Ellis:2016enq, Fuentes-Martin:2016uol, Zhang:2016pja, Ellis:2017jns}. Although this expanded CDE approach captures the majority of one-loop contributions in a matching calculation, it remains incomplete in the sense that no existing master formula accommodates all possible combinations of statistics and open derivatives in mixed diagrams. While such a master formula is likely to emerge soon, in the meantime the computation of the complete set of Wilson coefficients arising at one loop and dimension 6 in SMEFT for a specified, perturbative UV theory entails at least partial use of traditional Feynman diagram techniques. 

The potential relevance of various possible contributions to a one-loop matching calculation is a matter of some debate \cite{Henning:2016lyp}. Given a one-loop matching calculation at the scale $M$, Wilson coefficients $c_i(\mu)$ at a scale $\mu < M$ can be decomposed into four schematic contributions, namely 
\begin{eqnarray} \label{eq:cmu2}
c_i(\mu) &=& c_i^{(0)}(M) + c_{i,{\rm heavy}}^{(1)}(M) + c_{i,{\rm mixed}}^{(1)}(M) + \frac{1}{16 \pi^2} \sum_j \gamma_{ij} c_j^{(0)}(M) \log \frac{\mu}{M}
\end{eqnarray}
Here $c_i^{(0)}(M)$ represents tree-level contributions at the scale $M$; $c_{i,{\rm heavy}}^{(1)}(M)$ represents one-loop contributions at the scale $M$ from diagrams in which only heavy particles run in the loop; $c_{i,{\rm mixed}}^{(1)}(M)$ represents one-loop contributions at the scale $M$ from mixed diagrams in which both heavy and light particles run in the loop; and $c_j^{(0)}(M)$ represent tree-level contributions at the scale $M$ that feed into $c_i$ at one loop via running between $M$ and $\mu$. As we will see, while it is tempting to isolate some terms at the expense of others when performing a matching calculation, various combinations of these four terms may constitute the leading contribution to Wilson coefficients and, ultimately, to physical observables. This makes it worthwhile to compute the full set of one-loop contributions to Wilson coefficients in a given matching calculation.

In this work, we compute for the first time the full set of Wilson coefficients arising at one loop and dimension 6 in SMEFT for one of the most common EFT matching benchmarks: the real singlet scalar. The extension of the Standard Model by a real singlet scalar arises in a variety of motivated examples, including models addressing dark matter \cite{Silveira:1985rk, McDonald:1993ex, Burgess:2000yq, He:2008qm, Gonderinger:2009jp, Mambrini:2011ik}, baryogenesis \cite{Menon:2004wv, Huber:2006wf,Profumo:2007wc, Barger:2011vm, Espinosa:2011ax, Curtin:2014jma}, and the electroweak hierarchy problem \cite{Craig:2013xia, Curtin:2015bka}. The singlet scalar's signatures have been extensively studied at the LHC beginning with \cite{Datta:1997fx, BahatTreidel:2006kx, Barger:2006sk, OConnell:2006rsp, Barger:2007im,Barger:2008jx}, and its impact on the properties of the Higgs boson have made it a central target of recent EFT studies \cite{Henning:2014wua,deBlas:2014mba, Gorbahn:2015gxa, Brehmer:2015rna,Chiang:2015ura, delAguila:2016zcb, Buchalla:2016bse, Jiang:2016czg, Dawson:2017vgm, Corbett:2017ieo}. This strongly motivates improving the precision of the matching calculation between SMEFT and the Standard Model augmented by a real singlet scalar. The relative simplicity of this scenario also makes it an ideal setting for illustrating various subtleties associated with NLO matching and mapping.

We carry out the full one-loop matching calculation using a combination of CDE and Feynman diagram techniques in DR and $\overline{\rm MS}$. It bears emphasizing that the matching calculation is done off-shell, in order to extract the maximum amount of physical information, and reduced to operator coefficients in a redundant basis of dimension-6 SMEFT operators. For the sake of definiteness, we refer to this redundant basis generated by the matching procedure as the ``Green's basis.'' To convert operator coefficients in the Green's basis to a non-redundant basis, we first canonically normalize the fields (which typically accumulate finite wavefunction renormalization at one loop in our renormalization scheme) and then use equations of motion to eliminate redundant operators. In doing so we must additionally keep track of any tree-level shifts in Standard Model couplings associated with matching, which can influence one-loop Wilson coefficients when equations of motion are employed. Ultimately, for the sake of completeness and convenience we present the Wilson coefficients in both the Green's basis and the Warsaw basis.

The one-loop Wilson coefficients in this specific example also serve to illustrate a variety of general features in NLO matching to SMEFT at dimension 6. The one-loop structure of SMEFT possesses a variety of novel properties, including a surprising pattern of cancellations \cite{Alonso:2014rga} appearing in the one-loop renormalization of dimension-six operators \cite{Grojean:2013kd, Jenkins:2013zja, Jenkins:2013wua, Alonso:2013hga, Elias-Miro:2013gya, Elias-Miro:2013mua, Elias-Miro:2013eta}. These cancellations can be understood via non-supersymmetric non-renormalization theorems controlling the running of higher-dimension operators in four-dimensional quantum field theories \cite{Cheung:2015aba}. Unsurprisingly, the non-renormalization theorems necessarily extend to the logarithmic dependence of Wilson coefficients in a perturbative matching calculation, which we illustrate explicitly in the case of NLO matching for the singlet scalar. The non-renormalization theorems also have novel implications for the size of various contributions to Wilson coefficients at scales $\mu < M$. In particular, they signal the existence of cases in which mixed diagrams make the leading contribution to a Wilson coefficient or observable, creating a loophole in the general arguments of \cite{Henning:2016lyp} and highlighting the value of computing all contributions to matching at a given order even in the presence of arbitrary scale separation $\mu \ll M$.

Although our interest is primarily focused on complete one-loop matching in a specific UV completion, such data is most useful when coupled with a complete one-loop mapping to observables. To highlight some of the subtleties involved in mapping to observables at NLO, we compute the one-loop mapping to the $T$ parameter in the EFT obtained by integrating out the singlet scalar. While the $T$ parameter is not itself an observable, it nonetheless illustrates key aspects of NLO mapping and provides a non-trivial cross-check of our results.

The paper is organized as follows: In Section \ref{sec:tree} we set notation for the UV extension of the Standard Model by a real singlet scalar and lay the groundwork for matching to SMEFT at dimension 6. In particular, we define the germane dimension-6 SMEFT operators in both the Warsaw and Green's bases, and compute the tree-level Wilson coefficients in agreement with the literature. We then determine the full set of one-loop Wilson coefficients in Section \ref{sec:loop} using a mix of CDE and Feynman diagram techniques, presenting them in the Green's basis for completeness. These results highlight the novel interplay between non-supersymmetric non-renormalization theorems, the logarithmic dependence of Wilson coefficients, and the relevance of mixed diagrams, which we explore in Section \ref{sec:mixed}. In Section \ref{sec:obs} we highlight some of the subtleties involved in mapping to observables at NLO by mapping the singlet scalar EFT to the not-quite-observable $T$ parameter. Our conclusions are summarized in Section \ref{sec:conc}, and we present the full set of one-loop Wilson coefficients in the Warsaw basis in Appendix \ref{app:warsaw}.

\section{The Model and Tree-level Matching} \label{sec:tree}

Our primary goal is to compute the complete set of Wilson coefficients in the Standard Model EFT generated at dimension six, and up to one loop, by integrating out a heavy real singlet scalar $\phi$. Gauge invariance and Lorentz invariance permit this scalar to couple to the Standard Model exclusively through the Standard Model Higgs doublet $H$ at the renormalizable level. The admissible Lagrangian density for $\phi$ (up to tadpoles) is
\begin{equation} \label{eq:lang}
\mathcal{L} \supset \frac{1}{2}\left(\partial_\mu\phi\right)^2 - \frac{1}{2}M^2\phi^2 - A|H|^2\phi -\frac{1}{2}\kappa |H|^2 \phi^2 - \frac{1}{3!} \mu \phi^3 - \frac{1}{4!} \lambda_\phi \phi^4 -\frac{1}{2}\lambda_h |H|^4\, .
\end{equation}
We further define our conventions for the Standard Model couplings in Appendix \ref{app:warsaw}.

In general the interactions of $\phi$ can lead it to acquire a vacuum expectation value, particularly when the coupling to the Higgs is taken into account. In what follows we take (\ref{eq:lang}) to be the Lagrangian expanded about the vacuum of $\phi$, without requiring any specific relations between the couplings $A, \kappa, \mu,$ and $\lambda_\phi.$ The common-considered special case of a $\mathbb{Z}_2$ symmetry acting on $\phi$ corresponds to $A = \mu = 0.$ 

Integrating out $\phi$ at one loop will lead to nonzero Wilson coefficients for a redundant set of operators. Although we will present the full set of redundant coefficients in terms of operators in the Warsaw basis (Table \ref{tab:dim6ops}) and additional operators (Table \ref{tab:dim6greens}) that are distinguishable at the level of Green's functions, we will ultimately reduce to the Warsaw basis to consider their impact on $S$-matrix elements. This is done with an eye towards Section \ref{sec:mixed}, which explores features of the Wilson coefficients as they relate to the structure of the matrix of anomalous dimensions in SMEFT computed in the Warsaw basis.\footnote{Note that in a full Green's basis there are 4 operators that contain four Higgs fields and two derivatives. Between Tables \ref{tab:dim6ops} and \ref{tab:dim6greens} we define 5 such operators: this is purely for convenience in the matching calculation. $\CO_{H\Box}$ and $O_H$ are trivially related by integration by parts.}

\begin{table}[h]
\begin{center}{
\begin{tabular}{l l}
\hline \hline \vspace{-0.4cm} \\
$\CO_H = (H^\dag H)^3$ &    $\CO_{H W} = (H^\dagger H) W^{a}_{\mu \nu} W^{a\mu\nu}$  \\
$\CO_{H \Box} = (H^\dagger H)\Box(H^\dagger H)$ &  $\CO_{H B} = (H^\dagger H)B_{\mu\nu}B^{\mu\nu}$  \\
$\CO_{H D} = (H^\dagger D^\mu H)^*(H^\dagger D_\mu H)$ &$\CO_{H W\! B} = (H^\dagger \sigma^a H) W^{a}_{\mu\nu} B^{\mu\nu} $ \\   
$\CO_{uH} = (H^\dag H)(\bar q u \widetilde{H})$ & $\CO_{He} = (H^\dagger i
\overset{\leftrightarrow}{D}_\mu H)(\bar e \gamma^\mu e) $ \\
$\CO_{dH} = (H^\dag H)(\bar q d H) $ & $\CO^{(1)}_{H q} = (H^\dagger i
\overset{\leftrightarrow}{D}_\mu H)(\bar q \gamma^\mu q)$ \\
$\CO_{eH} = (H^\dag H)(\bar \ell e H) $ & $\CO^{(3)}_{H q} = (H^\dagger i
\overset{\leftrightarrow}{D} \ \!\!\!^{\, a}_\mu H)
(\bar q \gamma^\mu \sigma^a q)$\\
$\CO_{Hu} = (H^\dagger i
\overset{\leftrightarrow}{D}_\mu H)(\bar u \gamma^\mu u)$ & $\CO^{(1)}_{H \ell} = (H^\dagger i
\overset{\leftrightarrow}{D}_\mu H)(\bar\ell\gamma^\mu \ell)$ \\
$\CO_{Hd} = (H^\dagger i
\overset{\leftrightarrow}{D}_\mu H)(\bar d \gamma^\mu d) $ & $\CO^{(3)}_{H \ell} = (H^\dagger i
\overset{\leftrightarrow}{D} \ \!\!\!^{\, a}_\mu H)
(\bar\ell\gamma^\mu\sigma^a \ell)$ 
\\ \vspace{-0.4cm} \\
\hline \hline
\end{tabular}
\caption{The relevant subset of dimension-6 SMEFT operators in the Warsaw basis generated by integrating out a heavy singlet scalar at one loop. Here the $\sigma^a$ are the Pauli matrices, and $\overset{\leftrightarrow}{D} \ \!\!\!^{\, a}_\mu \equiv \overset{\leftrightarrow}{D}_\mu \sigma^a$.} \label{tab:dim6ops}
}\end{center}
\end{table}

\begin{table}[h]
\begin{center}
\begin{tabular}{l l}
\hline \hline \vspace{-0.4cm} \\
$O_{K4} =\Box H^\dagger \Box H$ &
$O_{HW} = i g (D^\mu H)^\dag \sigma^a (D^\nu H) W^a_{\mu \nu}$ \\
$O_R = H^\dagger H D^\mu H^\dagger D_\mu H$ &
$O_{HB} = i g' (D^\mu H)^\dag (D^\nu H) B_{\mu \nu}$ \\ 
$O_H = \frac{1}{2}(\partial_\mu (H^\dagger H))^2$ &
$O_W = \frac{i g}{2} (H^\dag \sigma^a \Dfbd H) D_\nu W^{\mu \nu, a}$    \\
$O_T = \frac{1}{2}(H^\dag \Dfbd H)^2 $ &
$O_B = \frac{i g'}{2} (H^\dag \Dfbd H) \partial_\nu B^{\mu \nu} $ \\
$O_{Hq}^{(1) \prime} = (H^\dagger H)(\bar q i \overset{\leftrightarrow}{\slashed{D}} q) $ &
$O_{Hu}^\prime = (H^\dagger H)(\bar u i \overset{\leftrightarrow}{\slashed{D}} u) $ \\
$O_{Hq}^{(3) \prime} = (H^\dagger \sigma^a H)(\bar q i \overset{\leftrightarrow}{\slashed{D}}  \ \!\!\!^{\, a} q) $ &
$O_{Hd}^\prime = (H^\dagger H)(\bar d i \overset{\leftrightarrow}{\slashed{D}} d) $ \\
$O_{H\ell}^{(1) \prime} = (H^\dagger H)(\bar \ell i \overset{\leftrightarrow}{\slashed{D}} \ell) $ &
$O_{He}^\prime = (H^\dagger H)(\bar e i \overset{\leftrightarrow}{\slashed{D}} e) $ \\
$O_{H\ell}^{(3) \prime} = (H^\dagger \sigma^a H)(\bar \ell i \overset{\leftrightarrow}{\slashed{D}}  \ \!\!\!^{\, a} \ell) $ &
\\ \vspace{-0.4cm} \\
\hline \hline
\end{tabular}
\caption{The additional dimension-6 SMEFT operators, produced by integrating out a real singlet scalar, that are in the Green's basis, but not the Warsaw basis. For consistency we use the sign and normalization conditions of \cite{Wells:2015uba}.\label{tab:dim6greens}}
\end{center}
\end{table}

The real singlet scalar is a particularly useful case for exploring the structure of matching at one loop because it generates only two Wilson coefficients at tree level, with a much larger set of Wilson coefficients generated at one loop. The two tree-level Wilson coefficients are simply obtained from the classical equation of motion, yielding the familiar results
\begin{eqnarray}
c^{(0)}_H &=& \frac{\mu A^3}{6 M^4}-\frac{\kappa A^2}{2M^2} , \label{eq:ch}\\
c^{(0)}_{H \Box} &=& -\frac{A^2}{2M^2} , \label{eq:chbox}
\end{eqnarray}
in addition to a contribution to the Higgs quartic of
\begin{eqnarray}
\delta \lambda_h &=& -\frac{A^2}{M^2} .
\end{eqnarray}
In the following, we define the Higgs quartic in the EFT as $\lambda \equiv \lambda_h+\delta \lambda_h$.

At the level of tree-level observables, the two dimension 6 pieces respectively lead to shifts in the Higgs self-coupling and the universal suppression of all Higgs couplings. If (\ref{eq:lang}) results from the spontaneous breaking of a $\mathbb{Z}_2$-symmetric potential, then $\mu, \kappa,$ and $A$ are related in such a way as to set $c_H = 0$ \cite{Gorbahn:2015gxa}. In the case of an unbroken $\mathbb{Z}_2$ symmetry acting on $\phi$, all tree-level effects vanish and the leading contributions to SMEFT first appear at one loop.

%\jmy{Besides these dimension-6 operators, the tree level matching also generate dimension-4 operators which will shift the SM couplings. For the singlet case, it is $\frac{A^2}{2M^2} |H|^4$. And it shifts the SM Higgs quartic by
%\begin{eqnarray}
%\lambda_h' = \lambda_h-\frac{A^2}{M^2}.
%\end{eqnarray}
%When we transform our results into Warsaw basis, it is this shifted $\lambda'$ that appears in the SM EOM. In the following, we use $\lambda$ for $\lambda'$ for simplicity}

\section{One-loop Matching} \label{sec:loop}

We now move to the Wilson coefficients at one loop, using a combination of the UOLEA and conventional Feynman diagram techniques.  Using the UOLEA we are able to obtain a subset of the one-loop Wilson coefficients, not all of which are in the Warsaw basis. Operators not in the Warsaw basis are denoted by $O_i$, rather than $\mathcal{O}_i$, and the corresponding coefficients are denoted by $\kappa_i$. As we will see, it is useful to retain possible logarithmic dependence of the Wilson coefficients on the renormalization scale $\mu$, although in matching to SMEFT one typically chooses $\mu = M$ to minimize these logarithms. 

The one-loop corrections to the tree-level Wilson coefficients for $c_H$ and $c_{H \Box}$ can be computed directly using the UOLEA, yielding
\begingroup
\allowdisplaybreaks
\begin{eqnarray}
16 \pi^2 \delta c_H &=& -\frac{\kappa ^3}{12}+\frac{-48 A^2 \kappa  \lambda +36 A^2 \lambda^2+22 A^2 \kappa ^2-2 A^2 \kappa  \lambda _{\phi }-A \kappa ^2 \mu }{4 M^2}\nonumber \\
&&+\frac{39 A^4 \kappa -3 A^4 \lambda _{\phi }+36A^3 \mu  \lambda-30 A^3 \kappa  \mu +2 A^3 \mu  \lambda _{\phi }}{6 M^4}\nonumber \\
&&+\frac{-8A^6-18 A^5 \mu +12 A^4 \mu ^2+7 A^3 \mu ^3}{12 M^6}+\left [\frac{36 A^2 \kappa  \lambda_h-36 A^2 \lambda^2-12 A^2 \kappa ^2+A^2 \kappa  \lambda _{\phi }}{4 M^2} \right. \nonumber \\
&&+\frac{-18 A^4 \kappa -A^3 \mu  \lambda+12 A^3 \kappa  \mu -A^3 \mu  \lambda _{\phi }+6 A^2 \kappa  \mu ^2}{12 M^4} \left. +\frac{2 A^5 \mu -3 A^3 \mu ^3}{4 M^6}\right] \log\frac{M^2}{\mu^2} , \label{eq:dch}\\
16 \pi^2 \delta c_{H \Box} &=&-\frac{\kappa ^2}{24}+\frac{-27 A^2 \lambda+34 A^2 \kappa -6 A^2 \lambda _{\phi }-5 A \kappa  \mu }{12 M^2}+\frac{26A^4-32 A^3 \mu -A^2 \mu ^2}{24 M^4}\nonumber \\
&&+\left[\frac{A^2 \lambda-4 A^2 \kappa +A^2 \lambda _{\phi }}{2 M^2}+\frac{A^2 \mu ^2-2 A^4}{2 M^4}\right]\log\frac{M^2}{\mu^2} , \label{eq:dchbox}
\end{eqnarray}
\endgroup
as were first computed in \cite{Ellis:2017jns}.

The Wilson coefficients readily extracted from the UOLEA  for operators first generated at one loop are
\begingroup
\allowdisplaybreaks
\begin{eqnarray}
16 \pi^2 \kappa_{K4}&=&\frac{1}{6}\frac{A^2}{M^2} ,\\
16 \pi^2 c_{HW} &=& \frac{1}{16} \frac{A^2}{M^2} g^2 ,\\
16 \pi^2 c_{HB} &=& \frac{1}{16} \frac{A^2}{M^2} g'^2 ,\\
16 \pi^2 c_{HWB} &=&  \frac{1}{8} \frac{A^2}{M^2} g g' ,\\
16 \pi^2 \kappa_R &=& \frac{A^3 \mu }{2 M^4}-\frac{3 A^2 \kappa -9 A^2 \lambda}{2 M^2}-\frac{A^2 \lambda}{M^2}\log\frac{M^2}{\mu^2},\\
16 \pi^2 \kappa_{HW} &=&  -\frac{1}{12} \frac{A^2}{M^2} , \\
16 \pi^2 \kappa_{HB} &=& -\frac{1}{12} \frac{A^2}{M^2}, \\
16 \pi^2 \kappa_W &=& \frac{1}{6} \left( -\frac{7}{3} +\log \frac{M^2}{\mu^2} \right) \frac{A^2}{M^2} ,\\
16 \pi^2 \kappa_B &=& \frac{1}{6} \left(- \frac{7}{3} + \log \frac{M^2}{\mu^2} \right) \frac{A^2}{M^2} .
\end{eqnarray}
\endgroup
At one loop there is also a correction to the Higgs kinetic term, $\delta Z_h D^\mu H^\dagger D_\mu H$, of
\begin{eqnarray}
16 \pi^2 \delta Z_h &=&\frac{1}{2}\frac{A^2}{M^2},\label{eq:dzh}
\end{eqnarray}
which, after the field redefinition $H\rightarrow (1- \frac12 \delta Z_h)H$, gives an additional one loop contribution to the coefficients of the tree level operators $\CO_H$ and $\CO_{H \Box}$:
\begin{eqnarray}
16\pi^2\delta c_H \big|_\text{shift} &=& -\frac{3A^2}{2M^2}c^{(0)}_H , \label{eq:dchshift} \\
16\pi^2\delta c_{H \Box} \big|_\text{shift} &=& -\frac{A^2}{M^2}c^{(0)}_{H \Box}. \label{eq:dchboxshift}
\end{eqnarray}

The UOLEA given in \cite{Ellis:2017jns} does not include results associated with open derivatives and mixed statistics. In our situation this means that contributions from loops with fermions and gauge bosons are not included, and so we match at the level of Feynman diagrams in these cases. Diagrams with fermions running in the loop give rise to the following additional Wilson coefficients
\begingroup
\allowdisplaybreaks
\begin{eqnarray}
16 \pi^2 c_{Hu} &=&  \left(\frac{5}{8}-\frac{1}{4}\log\frac{M^2}{\mu^2} \right) \frac{A^2}{M^2} y_u^\dagger y_u ,\\
16 \pi^2 \kappa_{Hu}^\prime &=&  \left(\frac{1}{8}-\frac{1}{4}\log\frac{M^2}{\mu^2} \right) \frac{A^2}{M^2} y_u^\dagger y_u ,\\
16 \pi^2 c_{Hd} &=&  \left(-\frac{5}{8}+\frac{1}{4}\log\frac{M^2}{\mu^2} \right) \frac{A^2}{M^2} y_d^\dagger y_d ,\\
16 \pi^2 \kappa_{Hd}^\prime &=&  \left(\frac{1}{8}-\frac{1}{4}\log\frac{M^2}{\mu^2} \right) \frac{A^2}{M^2} y_d^\dagger y_d ,\\
16 \pi^2 c_{He} &=&  \left(-\frac{5}{8}+\frac{1}{4}\log\frac{M^2}{\mu^2} \right) \frac{A^2}{M^2} y_e^\dagger y_e ,\\
16 \pi^2 \kappa_{He}^\prime &=&  \left(\frac{1}{8}-\frac{1}{4}\log\frac{M^2}{\mu^2} \right) \frac{A^2}{M^2} y_e^\dagger y_e ,\\
16 \pi^2 c_{uH} &=&   \left(1 - \log\frac{M^2}{\mu^2}\right) \frac{A^2}{M^2} y_u y_u^\dagger y_u ,\\ 
16 \pi^2 c_{dH} &=&   \left(1 - \log\frac{M^2}{\mu^2}\right) \frac{A^2}{M^2} y_d y_d^\dagger y_d ,\\ 
16 \pi^2 c_{eH} &=&   \left(1 - \log\frac{M^2}{\mu^2}\right) \frac{A^2}{M^2} y_e y_e^\dagger y_e ,\\ 
16 \pi^2 c_{Hq}^{(1)} &=&  \left( - \frac{5}{16} + \frac{1}{8} \log \frac{M^2}{\mu^2} \right) \frac{A^2}{M^2} ( y_u y_u^\dagger - y_d y_d^\dagger) ,\\
16 \pi^2 c_{Hq}^{(3)} &=&  \left( \frac{5}{16} - \frac{1}{8} \log \frac{M^2}{\mu^2} \right) \frac{A^2}{M^2} ( y_u y_u^\dagger + y_d y_d^\dagger), \\ 
16 \pi^2 \kappa_{Hq}^{(1)\prime} &=&  \left( \frac{1}{16} - \frac{1}{8} \log \frac{M^2}{\mu^2} \right) \frac{A^2}{M^2} ( y_u y_u^\dagger + y_d y_d^\dagger) ,\\
16 \pi^2 \kappa_{Hq}^{(3)\prime} &=&  \left( -\frac{1}{16} + \frac{1}{8} \log \frac{M^2}{\mu^2} \right) \frac{A^2}{M^2} ( y_u y_u^\dagger - y_d y_d^\dagger) ,\\ 
16 \pi^2 c_{Hl}^{(1)} &=&  \left( \frac{5}{16} - \frac{1}{8} \log \frac{M^2}{\mu^2} \right) \frac{A^2}{M^2} ( y_e y_e^\dagger) ,\\
16 \pi^2 c_{Hl}^{(3)} &=&  \left( \frac{5}{16} - \frac{1}{8} \log \frac{M^2}{\mu^2} \right) \frac{A^2}{M^2} ( y_e y_e^\dagger) ,\\ 
16 \pi^2 \kappa_{Hl}^{(1)\prime} &=&  \left( \frac{1}{16} - \frac{1}{8} \log \frac{M^2}{\mu^2} \right) \frac{A^2}{M^2} ( y_e y_e^\dagger) ,\\
16 \pi^2 \kappa_{Hl}^{(3)\prime} &=&  \left( \frac{1}{16} - \frac{1}{8} \log \frac{M^2}{\mu^2} \right) \frac{A^2}{M^2} ( y_e y_e^\dagger) ,
\end{eqnarray}
\endgroup
whereas diagrams with gauge bosons running in the loop give rise to
\begingroup
\allowdisplaybreaks
\begin{eqnarray}
16 \pi^2 \kappa_H&=&  \left (\frac{5}{8}-\frac{3}{4}\log\frac{M^2}{\mu^2} \right)\frac{A^2}{M^2} g^2 ,\\
16 \pi^2 \kappa_R&=&  \left(-\frac{5}{4}+\frac{3}{2}\log\frac{M^2}{\mu^2}\right) \frac{A^2}{M^2} g^2 , \\
16 \pi^2 \kappa_T&=& \left(\frac{5}{8}-\frac{3}{4}\log\frac{M^2}{\mu^2} \right) \frac{A^2}{M^2} g'^2 .\label{eq:t} 
\end{eqnarray}
\endgroup
%
%exclusively in terms of the non-Warsaw four Higgs operators.

This represents the complete set of Wilson coefficients generated at one loop and dimension six in SMEFT from integrating out a real singlet scalar. For convenience, the coefficients are reduced to coefficients entirely contained in the Warsaw basis in Appendix \ref{app:warsaw}. 

There are a variety of useful cross-checks for these results. The coefficients of terms logarithmic in the renormalization scale $\mu$ can be checked by appropriate comparison with known entries in the SMEFT matrix of anomalous dimensions, and agree with expressions appearing in the literature. Also of interest are Wilson coefficients that are entirely independent of the renormalization scale $\mu$ at one loop, whose properties will be discussed in the next section.

\section{Mixed Diagrams and Non-renormalization Theorems} \label{sec:mixed}

The majority of the Wilson coefficients computed above come from mixed diagrams, in which both heavy and light particles run in the loop. As emphasized in  \cite{Henning:2016lyp}, these mixed diagrams arise only when there is a nonzero tree-level Wilson coefficient, since the mixed diagram can be thought of as arising from the one-loop dressing of a tree-level diagram involving the exchange of a heavy particle. The tree-level Wilson coefficient then {\it typically} contributes to the same Wilson coefficient as the mixed diagram through one-loop running in SMEFT. Consequently, in exploring the practical relevance of mixed diagrams, \cite{Henning:2016lyp} identify four cases of interest for the calculation of Wilson coefficients:
\begin{enumerate}
\item There are no tree-level diagrams, in which case there are no mixed diagrams and the leading effects come from loop diagrams involving only heavy particles. In this case $c_{i,{\rm heavy}}^{(1)}(M)$ is the leading contribution in (\ref{eq:cmu2}).
\item There is a nonzero tree-level coefficient $c_j^{(0)}$ that feeds into $c_i$ at one loop through the matrix of anomalous dimensions, but there is also a tree-level contribution to $c_i$, i.e., nonzero $c_i^{(0)}$. In this case $c_i^{(0)}$ is the leading contribution in (\ref{eq:cmu2}).
\item There is a nonzero tree-level coefficient $c_j^{(0)}$ that feeds into $c_i$ at one loop through the matrix of anomalous dimensions, there is no tree-level contribution to $c_i$,  $c_i^{(0)} = 0$, and the separation of scales is large, $\mu \ll M$. Then there is a nonzero mixed coefficient $c_{i,{\rm mixed}}^{(1)}$, but the logarithmically-enhanced correction $\frac{1}{16 \pi^2} \gamma_{ij} c_j^{(0)}(M) \log \frac{\mu}{M}$ is the leading contribution in (\ref{eq:cmu2}).
\item There is a nonzero tree-level coefficient $c_j^{(0)}$ that feeds into $c_i$ at one loop through the matrix of anomalous dimensions, there is no tree-level contribution to $c_i$,  $c_i^{(0)} = 0$, and the separation of scales is not large, $\mu \sim M$. Then the mixed diagram $c_{i,{\rm mixed}}^{(1)}$ could be a leading contribution in (\ref{eq:cmu2}).
\end{enumerate} 
Only in the fourth case do mixed diagrams seem to play a practically significant role at one loop. This would seem to relegate mixed diagrams to a relatively limited set of cases, and precisely those cases ($\mu \sim M$) in which the validity of the dimension-6 SMEFT truncation is itself questionable. 

There is, however, a very interesting {\it fifth} possibility beyond those considered in  \cite{Henning:2016lyp}. While Wilson coefficients arising from mixed diagrams typically receive one-loop, logarithmically enhanced contributions from tree-level Wilson coefficients via running between $M$ and $\mu$ in SMEFT, this is not always the case. Famously, there are surprising zeroes in the SMEFT matrix of anomalous dimensions \cite{Alonso:2014rga} --- surprising in the sense that there appear to be one-loop diagrams that could contribute to the anomalous dimensions in question, but these diagrams turn out not to be logarithmically divergent. The pattern of these surprising zeroes was understood from the perspective of helicity selection rules in \cite{Cheung:2015aba}. Thus the fifth possibility:

\begin{enumerate}
\item[5.] There is a nonzero tree-level coefficient $c_j^{(0)}$, but it does not feed into $c_i$ at one loop through the matrix of anomalous dimensions because $\gamma_{ij} = 0$ despite the existence of relevant one-loop diagrams. Then the mixed diagram $c_{i,{\rm mixed}}^{(1)}$ is a leading contribution in (\ref{eq:cmu2}), regardless of the separation of scales.\footnote{Note, however, that in the one-loop mapping to observables, there are contributions of size comparable to $c_{i,{\rm mixed}}^{(1)}$ from the finite one-loop graph containing an insertion of $c_j^{(0)}$.}
\end{enumerate}

The fifth possibility defines an interesting set of cases in which mixed diagrams comprise the leading contribution to Wilson coefficients, intimately connected to the pattern of surprising zeroes in the SMEFT matrix of anomalous dimensions. Moreover, these cases are of more than academic interest, in that the Wilson coefficients arising predominantly from mixed diagrams are often also the leading contributions to specific observables. 

This is illustrated nicely in the case of the Wilson coefficients arising from integrating out a real singlet scalar. Only two coefficients are generated at tree-level: $c_{H \Box}$, the coefficient of the two-derivative, four-scalar operator $\mathcal{O}_{H \Box}$ (schematically in the category of $\phi^4 D^2$ operators in the notation of \cite{Cheung:2015aba}), and $c_H$, the coefficient of the six-scalar operator $\mathcal{O}_H$ (schematically the sole member of the category of $\phi^6$ operators). From the helicity selection rule argument in \cite{Cheung:2015aba}, we expect SMEFT operators in the $\phi^4 D^2$ category to renormalize operators in the categories $\psi^2 \phi^3$ and $\bar \psi^2 \phi^3$ (two-fermion, three-scalar operators); $\bar \psi \psi \phi^2 D$ (one-derivative, two-fermion, two-scalar operators); $\phi^4 D^2$; and $\phi^6$ at one loop. We do not expect operators in $\phi^4 D^2$ to renormalize a host of other operators (three-field-strength operators $F^3, \bar F^3$; one-field-strength, two-fermion, one-scalar operators $F \psi^2 \phi, \bar F \bar \psi^2 \phi$; and four-fermion operators $\psi^4, \bar \psi^4, \bar \psi^2 \psi^2$) because there are no diagrams. More interestingly, there {\it are} diagrams that could potentially allow $\phi^4 D^2$ operators to renormalize two-field-strength, two-scalar operators $F^2 \phi^2$ and $\bar F^2 \phi^2$, but these diagrams do not contain logarithmic divergences on account of Standard Model helicity selection rules. For SMEFT operators in the category $\phi^6$, the situation is far simpler: $\phi^6$ operators can only renormalize operators in the same category at one loop, and there are no diagrams that could possibly allow the renormalization of operators from other categories.

In the case of the singlet scalar, we thus expect that one-loop mixed diagrams contributing to Wilson coefficients for operators in the $F^2 \phi^2$ and $\bar F^2 \phi^2$ categories provide the {\it leading} contribution to these coefficients at a scale $\mu \leq M$. And, indeed, this is what we observe. Working in the Warsaw basis, the leading contributions to the Wilson coefficients for $\mathcal{O}_{HW}, \mathcal{O}_{HB},$ and $\mathcal{O}_{HWB}$ at the scale $M$ come from mixed diagrams. The only operators with tree-level Wilson coefficients are $\mathcal{O}_H$ and $\mathcal{O}_{H \Box}$, which do not renormalize $\mathcal{O}_{HW}, \mathcal{O}_{HB},$ or $\mathcal{O}_{HWB}$ at one loop, and so the mixed diagram contribution to $\mathcal{O}_{HW}, \mathcal{O}_{HB},$ and $\mathcal{O}_{HWB}$ is the dominant one-loop contribution at {\it any} scale $\mu < M$.

This pattern of renormalization in SMEFT also has interesting implications for the structure of the Wilson coefficients themselves. If a one-loop mixed diagram gives the leading contribution $c_{i,{\rm mixed}}^{(1)}$ to the Wilson coefficient of the operator $\mathcal{O}_i$, then the Wilson coefficient $c_{i,{\rm mixed}}^{(1)}$ must be independent of the renormalization scale $\mu$ at the same order. This is not the case, of course, for one-loop Wilson coefficients of operators with either a corresponding tree-level contribution {\it or} a one-loop contribution from a different tree-level contribution via the matrix of anomalous dimensions. Thus the surprising zeroes in the matrix of anomalous dimensions necessarily signal the existence of one-loop Wilson coefficients that are independent of the renormalization scale $\mu$. Indeed, this is what is observed for the real singlet scalar: the Wilson coefficients $\mathcal{O}_{HW}, \mathcal{O}_{HB},$ and $\mathcal{O}_{HWB}$ are independent of $\mu$ at one loop. The interplay between the SMEFT matrix of anomalous dimensions and scale dependence of Wilson coefficients is illustrated schematically in Table~\ref{tab:zeroes}, and highlights the sense in which the renormalization of operators in SMEFT also governs the properties of Wilson coefficients in a perturbative matching calculation.

In rare cases, the leading contribution to a given Wilson coefficient comes from a one-loop mixed diagram with only one light particle in the loop. In such a case, the Wilson coefficient is necessarily independent of $\mu$, for there will be no possible diagram through which any tree level coefficient may renormalize it. This is the case for the $\bar{\psi}^2 \psi^2$ operators ($\mathcal{O}_{2y}$ in Appendix \ref{app:warsaw}) in the singlet example, which arise from $O_{K4}$ after applying EOM.

\begin{table*}[t]
{\small	$\begin{array} {|*2c|  *5c| *5c| *4c| } \hline
	& & F^{3} & F^{2} \phi^2 & F \psi^2 \phi & \psi^4 & 
	\psi^2 \phi^3 & 
	\bar{F}^3 & \bar{F}^2 \phi^2 & \bar{F}\bar{\psi}^{2} \phi & \bar{\psi}^4 & \bar{\psi}^{2} \phi^3 & \bar{\psi}^2 \psi^2 & \bar{\psi}\psi \phi^2 D & \phi^4 D^2 & \phi^6  \\
	& (w,\bar{w}) & (0,6) & (2,6) & (2,6) & (2,6) & (4,6) & (6,0) & (6,2) & (6,2) & (6,2) &
	(6,4) & (4,4) & (4,4) & (4,4) & (6,6) \\ \hline
	\rowcolor{zero2}
	\cw\phi^4 D^2 & \cw(4,4) & \nda & \czb \dag & \nda & \nda & \cw \checkmark & \nda & \czb \dag & \nda & \nda & \cw \checkmark & \cw \nda \dag & \cw \checkmark & \cw \checkmark & \cw \checkmark \\
	\rowcolor{zero2}
	\cw\phi^6 & \cw(6,6) & \nda & \nda & \nda & \nda & \nda & \nda & \nda  & \nda & \nda & \nda & \nda & \nda & \nda & \cw \checkmark \\ \hline
	\end{array}$ }
\caption{Interplay between the SMEFT matrix of anomalous dimensions and the renormalization scale dependence of Wilson coefficients for the real singlet scalar. The rows correspond to operator categories with tree-level Wilson coefficients for the real singlet scalar, which may or may not renormalize the operator category indicated in each column at one loop. The $(w,\bar w)$ refer to the `holomorphic weights' defined in \cite{Cheung:2015aba}. The shaded entries indicate cases where the operator category in question is not renormalized by one of the tree-level operators due to the non-renormalization theorems of \cite{Cheung:2015aba}, in agreement with~\cite{Alonso:2014rga}. The $\times$ labels indicate cases where the operator in question is not renormalized by one of the tree-level operators because there are no diagrams~\cite{Jenkins:2013sda}. These general results are consistent with the logarithmic renormalization scale dependence of Wilson coefficients for the singlet scalar: The $\checkmark$ labels denote operator categories with one-loop Wilson coefficients that depend logarithmically on the renormalization scale, while the $\dag$ labels denote operator categories with one-loop Wilson coefficients that are strictly independent of the renormalization scale. The entry marked `$\nda \dag$' refers to the case where there is no one-loop diagram in the EFT; however, there may still be a finite one-loop contribution to the operator coefficient from a mixed diagram in the UV theory, cf.~our discussion at the end of Section \ref{sec:mixed}.}
\label{tab:zeroes}
\end{table*}

\section{Mapping to Observables} \label{sec:obs}

Of course, computing the full set of one-loop Wilson coefficients at the matching scale is but one step towards the calculation of observables at one loop. The one-loop SMEFT matrix of anomalous dimensions allows these Wilson coefficients to be run to some lower scale $\mu$ at which observables are to be computed, but it then remains to map onto the appropriate observables in the EFT at one loop as well. Here we illustrate some of the subtleties associated with the one-loop mapping to observables by considering the real singlet scalar contributions to the $T$ parameter.\footnote{While oblique parameters are not themselves observables, the $T$ parameter is nonetheless a useful case study for mapping at one loop in SMEFT.} This may be computed directly at one loop in the full theory after electroweak symmetry breaking in the mass eigenbasis, giving \cite{Barger:2007im}
 \begin{equation} \label{eq:fullt}
\Delta T_{\rm full}=-\frac{3}{16\pi s_W^2}\sin^2\theta \left( \left[\frac{1}{c_W^2}\frac{m_2^2}{m_2^2-M_Z^2}\log\frac{m_2^2}{M_Z^2}-\frac{m_2^2}{m_2^2-M_W^2}\log\frac{m_2^2}{M_W^2} \right]-m_2\rightarrow m_1 \right) \, .
\end{equation}
Here $\theta$ is the mixing angle between the singlet and the Higgs after electroweak symmetry breaking and $m_2, m_1$ are the mass eigenvalues; we identify $m_1=m_h$ as the SM-like Higgs mass. We expect agreement with the dimension-6 EFT result only to combined second order in the ratios $x=v/M, y=m_h/M$. Expanding (\ref{eq:fullt}) to this order gives
 \begin{equation}
\Delta T_{\rm full}=\left( \frac{3A^2(g'^2+g^2)\log y}{8\pi g^2 M^2 }+\mathcal{O}(y^2) \right) x^2+\mathcal{O}(x^3) \, .
\end{equation}
In the EFT, it is helpful to compute $\Delta T$ in a particular non-redundant operator basis. In the Warsaw basis, the $T$ parameter is simply related to our Wilson coefficients at tree level via
\begin{equation}
\Delta T_{\rm EFT}^{(0)} = \frac{v^2}{\alpha M^2} \left( -\frac{1}{2} c_{HD} + \frac{g'^2}{2} c_{HJB} \right),
\label{eq:efft}
\end{equation}
where $c_{HJB}$ is the coefficient of the operator combination \cite{Wells:2015uba}\footnote{$Y_f$ is the hypercharge of the fermion $f$.}
\begin{eqnarray}
\mathcal{O}_{HJB} &\equiv& \frac{i g'}{2} (H^\dag \Dfbd H) J_B^\mu \\
&=& \frac{1}{2} g'^2 \left( Y_q [\mathcal{O}_{Hq}^{(1)}]_{ii} + Y_l [\mathcal{O}_{Hl}^{(1)}]_{ii} +  Y_u [\mathcal{O}_{Hu}]_{ii}+Y_d [\mathcal{O}_{Hd}]_{ii} + Y_e [\mathcal{O}_{He}]_{ii} \right) \, .
\end{eqnarray}
In practice, $c_{HJB}$ appears in $\Delta T_{\rm EFT}$ upon elimination of the operator $O_B$, which also gives a compensating contribution to the coefficient $c_{HD}$. Matters are somewhat simpler in e.g.~the SILH basis, where $O_T$ is retained, and the contribution to $\Delta T_{\rm EFT}$ is just proportional to $\kappa_T$ in (\ref{eq:t}). In either case, matching at $\mu = M$, we obtain
\begin{equation}
\Delta T_{\rm EFT}^{(0)} = \frac{5}{32} \frac{A^2 (g'^2 + g^2)}{\pi g^2 M^2} \frac{v^2}{M^2} \, .
\end{equation}

However, care must be taken at one loop, where there are two additional contributions from loop-level processes involving an insertion of the tree-level coefficient $c_{H \Box}$. The first is the one loop running of Wilson coefficients between $M$ and lower scales proportional to $c_{H \Box}$. In principle, the SMEFT RGEs could be used in this case to resum the potentially large logarithm between $M$ and $m_h$. In practice, here we are just interested in agreement with the fixed-order calculation in the full theory, so we could equally capture the leading logarithmic behavior by simply leaving the renormalization scale $\mu$ unfixed in the expression for $\kappa_T$ in (\ref{eq:t}). Including this contribution gives 
\begin{equation}
\Delta T_{\rm EFT}^{(1a)} =\left( \frac{5}{32} - \frac{3}{8} \log \frac{M}{\mu}\right) \frac{A^2 (g'^2 + g^2)}{\pi g^2 M^2} x^2 \, .
\end{equation}

The second effect involves one-loop corrections to (\ref{eq:efft}) proportional to tree-level Wilson coefficients, namely $c_{H\Box}$. The effect of this operator can be most easily seen by going to unitary gauge. In the broken phase, this 
includes a term $-\frac{v^2}{M^2} c_{H\Box} (\partial h)^2$. After the redefinition of the Higgs field $h\rightarrow(1+\frac{v^2}{M^2} c_{H\Box})h$, the contribution to $T$ from the Standard Model sector is modified. In unitary gauge, the modification is to the vacuum polarization diagrams involving two $hVV$ vertices that are changed by the field redefinition.  The leading contribution from this part is
\begin{eqnarray}
\Delta T_{\rm EFT}^{(1b)}&=&-v^2 A^2/M^4 T_{\rm SM}\nonumber\\
&=&\frac{A^2(g'^2+g^2)}{\pi g^2 M^2 } \left(-\frac{5}{32}+\frac{3}{8}\log\frac{m_h}{\mu}+\mathcal{O}(y^3) \right)x^2+\mathcal{O}(x^3) \, .
\end{eqnarray}
Adding these corrections, the combined expression in the EFT at one loop is 
 \begin{equation}
\Delta T_{\rm EFT}=\left(\frac{3A^2(g'^2+g^2)\log y}{8\pi g^2 M^2 }+\mathcal{O}(y^3) \right)x^2+\mathcal{O}(x^3) \, ,
\end{equation}
in perfect agreement with the full theory to the appropriate order in $v/M$. It bears emphasizing that simply applying tree-level relations between SMEFT operators and precision electroweak observables does not yield a consistent result if the Wilson coefficients appearing in these relations are generated at one loop, while other Wilson coefficients arise at tree level.  This illustrates some of the subtleties involved in making use of a full NLO matching calculation in mapping to observables at NLO.

\section{Conclusion} \label{sec:conc}

The precision calculation of Wilson coefficients in matching calculations between perturbative UV completions of the Standard Model and the Standard Model EFT is a key step in maximizing the utility of the EFT framework. In this work we have presented, for the first time, the complete one-loop matching (in DR and $\overline{\rm MS}$) between the real scalar extension of the Standard Model and the dimension-6 SMEFT. Our calculation uses a combination of UOLEA and Feynman diagram techniques. The complete one-loop calculation involves a variety of subtleties, including the effects of both tree-level shifts to parameters in the Standard Model part of the SMEFT Lagrangian and one-loop wavefunction renormalization of the Higgs in the unbroken phase. In the interest of completeness, we have presented results for Wilson coefficients in both the redundant basis (``Green's basis'') generated by the matching calculation, and the non-redundant Warsaw basis.

Our results provide a key part of the information necessary for a complete one-loop EFT calculation of observables arising from the singlet scalar extension of the Standard Model. In particular, they may be fruitfully combined with the complete dimension-6 SMEFT matrix of anomalous dimensions and the one-loop mapping of SMEFT onto particular observables to generate partially resummed one-loop predictions for Standard Model deviations in this scenario. 

Apart from their immediate relevance to the precision calculation of observables in singlet extensions of the Standard Model, our results also illustrate a variety of general features of one-loop matching. In particular, we have explored the interplay between non-supersymmetric non-renormalization theorems, the logarithmic (in)dependence of Wilson coefficients, and the relevance of mixed diagrams in theories with large separation of scales. In addition, we have highlighted some of the subtleties involved in computing observables at next-to-leading order in SMEFT by mapping our results to the $T$ parameter at one loop, finding agreement with a one loop calculation in the full theory.

\acknowledgments

We thank Seth Koren and Zhengkang Zhang for useful conversations. MJ is supported by China Scholarships Council ( NO. 201706190097). NC and DS are supported in part by the US Department of Energy under the Early Career Award DE-SC0014129. YYL is supported by the Hong Kong PhD Fellowship Scheme (HKPFS) and thanks the Kavli Institute for Theoretical Physics for the award of a graduate visiting fellowship, provided through Simons Foundation Grant No.~216179 and Gordon and Betty Moore Foundation Grant No.~4310. We also thank the Kavli Institute of Theoretical Physics for hospitality during the inception of this work, and corresponding support from the National Science Foundation under Grant No. NSF PHY-1748958.

\appendix

\section{Warsaw Basis} \label{app:warsaw}

We define the relevant dimension 4 coefficients of the Standard Model \emph{in the EFT} via the lagrangian
\begin{eqnarray}
\mathcal{L} &=& \sum_{f=q,u,d,l,e} \bar{f} i \slashed{D} f + D_\mu H^\dagger D^\mu H + \mu_h^2 H^\dagger H - \frac14 B_{\mu\nu} B^{\mu\nu} - \frac14 W^a_{\mu\nu} W^{a \mu\nu} - \frac14 G^A_{\mu\nu} G^{A \mu\nu} \nonumber \\
&& - \frac12 \lambda (H^\dagger H)^2 - \left([y_u]_{ij} \bar q_i \tilde H u_j + [y_d]_{ij} \bar q_i H d_j + [y_e]_{ij} \bar \ell_i H e_j + \mathrm{h.c.} \right),
\end{eqnarray}
where the covariant derivative $D_\mu q = (\partial_\mu - \frac12 i g_s \lambda^A G^A_\mu - \frac12 i g \sigma^a W^a_\mu - \frac16 i g^\prime B_\mu ) q$ for the hypercharge $Y_q = \frac16$ left-handed quark, expressed in terms of the Gellmann $\lambda^A$ and Pauli $\sigma^a$ matrices (the covariant derivative is defined \emph{mutatis mutandis} for the other matter fields). We stress that the EFT values of these coefficients are shifted relative to those of the UV theory by integrating out the heavy scalar; at tree level,
\begin{eqnarray}
\lambda&=&\lambda_h -\frac{A^2}{M^2} .
\end{eqnarray}
Excepting the correction to the wavefunction renormalization of the Higgs in (\ref{eq:dzh}), the one loop shifts of the dimension 4 parameters of the EFT relative to their UV counterparts are unnecessary for the purposes of the one-loop matching and mapping to dimension 6 observables in the EFT.

Linear combinations of operators that are proportional to marginal equations of motion (in the EFT) do not contribute to $S$-matrix elements at dimension 6 order. Combining such EOM relations with IBP and Fierz relations, we arrive at the following operator identities (we only retain the dimension 6 parts):
\begingroup
\allowdisplaybreaks
\begin{eqnarray}
O_{K4} &=&\lambda^2\mathcal{O}_{H}+\lambda\left( [y_u]_{ij} [\mathcal{O}_{uH}]_{ij} + [y_d]_{ij} [\mathcal{O}_{dH}]_{ij} + [y_e]_{ij} [\mathcal{O}_{eH}]_{ij} + \mathrm{h.c.} \right)+\mathcal{O}_{2y},\\
O_H &=& -\frac{1}{2}\mathcal{O}_{H\Box},\\
O_R &=& \lambda \mathcal{O}_H+\frac{1}{2}\mathcal{O}_{H\Box}+ \left( \frac{1}{2} [y_u]_{ij} [\mathcal{O}_{uH}]_{ij} + \frac{1}{2} [y_d]_{ij} [\mathcal{O}_{dH}]_{ij} + \frac{1}{2} [y_e]_{ij} [\mathcal{O}_{eH}]_{ij} + \mathrm{h.c.} \right) ,\\
O_T &=& -2\mathcal{O}_{HD}-\frac{1}{2}\mathcal{O}_{H\Box} ,\\
O_{HB} &=& O_{B} - \frac{g'^2}{4} \mathcal{O}_{HB} - \frac{gg'}{4} \mathcal{O}_{HWB} ,\\
O_{HW} &=&  O_{W} -  \frac{g^2}{4} \mathcal{O}_{HW} - \frac{g g'}{4} \mathcal{O}_{HWB} ,\\
O_B &=& g'^2 \mathcal{O}_{HD} + \frac{1}{4}g'^2 \mathcal{O}_{H\Box}+ \frac{1}{2} g'^2 Y_q [\mathcal{O}_{Hq}^{(1)}]_{ii} + \frac{1}{2} g'^2 Y_l [\mathcal{O}_{Hl}^{(1)}]_{ii} + \frac{1}{2} g'^2 Y_u [\mathcal{O}_{Hu}]_{ii} \nonumber \\ && + \frac{1}{2} g'^2 Y_d [\mathcal{O}_{Hd}]_{ii} + \frac{1}{2} g'^2 Y_e [\mathcal{O}_{He}]_{ii} ,  \\
O_W &=& \frac{3}{4} g^2 \mathcal{O}_{H\Box}+ g^2\lambda \mathcal{O}_H  + \frac{1}{4} g^2 [\mathcal{O}_{Hq}^{(3)}]_{ii} + \frac{1}{4} g^2 [\mathcal{O}_{Hl}^{(3)}]_{ii} \nonumber \\ 
 && + \left( \frac{1}{2} g^2 [y_u]_{ij} [\mathcal{O}_{uH}]_{ij} + \frac{1}{2} g^2 [y_d]_{ij} [\mathcal{O}_{dH}]_{ij} +\frac{1}{2} g^2 [y_e]_{ij} [\mathcal{O}_{eH}]_{ij}+\mathrm{h.c.} \right), \\ \relax 
[O_{Hq}^{(1) \prime}]_{ij} &=& [y_u]_{jk} [\CO_{uH}]_{ik} + [y_d]_{jk} [\CO_{dH}]_{ik} + \mathrm{h.c.},\\ \relax
[O_{Hq}^{(3) \prime}]_{ij} &=& -[y_u]_{jk} [\CO_{uH}]_{ik} + [y_d]_{jk} [\CO_{dH}]_{ik} + \mathrm{h.c.},\\ \relax 
[O_{Hl}^{(1) \prime}]_{ij} &=& [y_e]_{jk} [\CO_{eH}]_{ik} + \mathrm{h.c.}, \\ \relax
[O_{Hl}^{(3) \prime}]_{ij} &=& [y_e]_{jk} [\CO_{eH}]_{ik} + \mathrm{h.c.}, \\ \relax
[O_{Hu}^\prime]_{ij} &=& [y_u]_{ki} [\CO_{uH}]_{kj} + \mathrm{h.c.}, \\ \relax
[O_{Hd}^\prime]_{ij} &=& [y_d]_{ki} [\CO_{dH}]_{kj} + \mathrm{h.c.}, \\ \relax
[O_{He}^\prime]_{ij} &=& [y_e]_{ki} [\CO_{eH}]_{kj} + \mathrm{h.c.}, 
\end{eqnarray}
\endgroup
where $Y_f$ is the hypercharge of the fermion $f$, and $\mathcal{O}_{2y}=|\bar{u}y_u^{\dagger}q_{\beta}\epsilon^{\beta \alpha}+\bar{q}^{\alpha} y_d d+\bar{l}^{\alpha}y_e e|^2$ \cite{Wells:2015uba} is a combination of the $\bar{\psi}^2\psi^2$ type operators in the Warsaw basis (we define the Levi-Civita epsilon to satisfy $\epsilon^{12}=+1$). We therefore reduce the redundant set of Wilson coefficients obtained by integrating out $\phi$ exclusively into coefficients of operators in the Warsaw basis:
\begingroup
\allowdisplaybreaks
\begin{eqnarray}
16 \pi^2  c_H &=& \frac{\lambda}{18} \left(84\lambda-31 g^2-27\kappa+\frac{9A\mu}{M^2}+ (30 g^2-18\lambda) \log \frac{M^2}{\mu^2} \right) \frac{A^2}{M^2} \nonumber\\
&+& 16 \pi^2 (\delta c_H+\delta c_H|_{\text{shift}}),\\
16 \pi^2  c_{H \Box} &=& \frac{1}{72} \left(81\lambda-93g^2-31g'^2-54\kappa+\frac{18A\mu}{M^2}+ (90g^2+30g'^2-36\lambda) \log \frac{M^2}{\mu^2} \right) \frac{A^2}{M^2}\nonumber\\
&+& 16 \pi^2 (\delta c_{H\Box}+\delta c_{H\Box}|_{\text{shift}}),\\
16 \pi^2  c_{HD} &=& \frac{1}{18} g'^2 \left(-31 + 30 \log \frac{M^2}{\mu^2} \right) \frac{A^2}{M^2},\\
16 \pi^2  c_{HW} &=& \frac{1}{12} g^2 \frac{A^2}{M^2},\\
16 \pi^2  c_{HB} &=& \frac{1}{12} g'^2 \frac{A^2}{M^2},\\
16 \pi^2  c_{HWB} &=&  \frac{1}{6} g g' \frac{A^2}{M^2},\\
16 \pi^2  c_{Hu} &=& \frac{1}{216} \left( -34 g'^2 + 135 y_u^\dagger y_u + (12 g'^2 - 54 y_u^\dagger y_u) \log \frac{M^2}{\mu^2} \right) \frac{A^2}{M^2} ,\\
16 \pi^2  c_{Hd} &=& \frac{1}{216} \left(17 g'^2 - 135 y_d^\dagger y_d - (6 g'^2 - 54 y_d^\dagger y_d) \log \frac{M^2}{\mu^2} \right) \frac{A^2}{M^2} ,\\
16 \pi^2  c_{He} &=& \frac{1}{72} \left( 17 g'^2 - 45 y_e^\dagger y_e - (6 g'^2 - 18 y_e^\dagger y_e) \log \frac{M^2}{\mu^2} \right) \frac{A^2}{M^2} ,\\
16 \pi^2  c_{uH} &=&\frac{1}{36}  y_u \Big( 45 y_u^\dagger y_u - 31 g^2 -27\kappa+ 87\lambda + 9\frac{A\mu}{M^2} \nonumber \\
&& \qquad \qquad \qquad +(30 g^2 -54 y_u^\dagger y_u - 18 \lambda)\log \frac{M^2}{\mu^2} \Big) \frac{A^2}{M^2},\\
16 \pi^2  c_{dH} &=&\frac{1}{36}  y_d \Big( 45 y_d^\dagger y_d - 31 g^2 -27\kappa+ 87\lambda + 9\frac{A\mu}{M^2} \nonumber \\
&& \qquad \qquad \qquad +(30 g^2 -54 y_d^\dagger y_d - 18 \lambda)\log \frac{M^2}{\mu^2} \Big) \frac{A^2}{M^2},\\
16 \pi^2  c_{eH} &=&\frac{1}{36}  y_e \Big( 45 y_e^\dagger y_e - 31 g^2 -27\kappa+ 87\lambda + 9\frac{A\mu}{M^2} \nonumber \\
&& \qquad \qquad \qquad +(30 g^2 -54 y_e^\dagger y_e - 18 \lambda)\log \frac{M^2}{\mu^2} \Big) \frac{A^2}{M^2},\\
16 \pi^2  c_{Hq}^{(1)} &=& \frac{1}{432} \Big( -17 g'^2 - 135 (y_u y_u^\dagger - y_d y_d^\dagger) \nonumber \\
&& \qquad \qquad \qquad + (6 g'^2 + 54 (y_u y_u^\dagger - y_d y_d^\dagger) ) \log \frac{M^2}{\mu^2} \Big) \frac{A^2}{M^2},\\
16 \pi^2  c_{Hq}^{(3)} &=& \frac{1}{144} \Big( -17 g^2 + 45 (y_u y_u^\dagger + y_d y_d^\dagger) \nonumber \\
&& \qquad \qquad \qquad + (6 g^2 - 18 (y_u y_u^\dagger + y_d y_d^\dagger) ) \log \frac{M^2}{\mu^2} \Big) \frac{A^2}{M^2},\\
16 \pi^2  c_{Hl}^{(1)} &=& \frac{1}{144} \left( 17 g'^2 + 45 y_e y_e^\dagger - (6 g'^2 +18 y_e y_e^\dagger) \log \frac{M^2}{\mu^2} \right) \frac{A^2}{M^2},\\
16 \pi^2  c_{Hl}^{(3)} &=& \frac{1}{144} \left( -17 g^2 + 45 y_e y_e^\dagger + (6 g^2 - 18 y_e y_e^\dagger) \log \frac{M^2}{\mu^2} \right) \frac{A^2}{M^2}, \\
16 \pi^2  c_{2y} &=& \frac{1}{6} \frac{A^2}{M^2} .
\end{eqnarray}
\endgroup
These one-loop coefficients are in addition to the tree-level contributions to $c_H$ and $c_{H\Box}$ detailed in (\ref{eq:ch}) and (\ref{eq:chbox}), respectively. For the sake of conciseness, here $\delta c_H$, $\delta c_{H \Box}$, $\delta c_H \big|_\text{shift}$ and $\delta c_{H \Box} \big|_\text{shift}$ refer to the one-loop expressions in (\ref{eq:dch}), (\ref{eq:dchbox}), (\ref{eq:dchshift}), and (\ref{eq:dchboxshift}). In the fermionic operators' coefficients, the absence of any flavor structure in a given term (i.e.~the absence of Yukawa matrices) should be read as a Kronecker delta $\delta_{ij}$ in the flavor indices.

\bibliography{singletbib}

\end{document}